\documentclass[runningheads]{llncs}
\usepackage{graphicx,amsmath}
\usepackage[T1]{fontenc}

\graphicspath{{./}{./figs/}}

\usepackage{nameref,hyperref}
\usepackage[table]{xcolor}
\hypersetup{
    colorlinks = true,                    
    citecolor = {blue},
    linkcolor = {teal},
           }

\begin{document}

\title{Self-organized attractoring in locomoting animals and robots: an emerging field}
\titlerunning{Self-organized attractoring in locomoting animals and robots}
\author{Bulcs{\'u} S\'andor\inst{1} 
\and Claudius Gros\inst{2}}
\authorrunning{B. S\'andor et al.}
%
\institute{Department of Physics, Babes-Bolyai University, Cluj-Napoca, Romania 
\email{bulcsu.sandor[[there]]ubbcluj.ro} \and
Institute for Theoretical Physics, Goethe University Frankfurt, \\
Frankfurt am Main, Germany\\
\email{gros[[here]]itp.uni-frankfurt.de}}
\maketitle              
\begin{abstract}        
Locomotion may be induced on three levels. On a 
classical level, actuators and limbs follow the 
sequence of open-loop top-down control signals 
they receive. Limbs may move alternatively on 
their own, which implies that interlimb coordination 
must be mediated either by the body or via decentralized
inter-limb signaling. In this case, when embodiment is 
present, two types of controllers are conceivable for the 
actuators of the limbs, local pacemaker circuits and control 
principles based on self-organized embodiment. The latter,
self-organized control, is based on limit cycles and chaotic 
attractors that emerge within the feedback loop composed of 
controller, body, and environment. For this to happen, the 
sensorimotor loop must be locally closed, e.g.\ via propriosensation.
Here we review the progress made within the framework of 
self-organized embodiment, with a particular focus on the 
concept of attractoring. This concept characterizes situations 
when sets of attractors combining discrete and continuous 
spectra are available as motor primitives for higher-order 
control schemes, such as kick control. In particular, we show 
that a simple generative principle allows for the robust 
formulation of self-organized embodiment. Based on the recurrent 
alternation between measuring the actual status of an actuator
and providing a target for the actuator to achieve in the next
step, we find that the mechanism leads to compliant locomotion 
for a range of simulated and real-world robots, which include 
barrel- and sphere-shaped agents, as well as wheeled and legged 
robots.

\end{abstract}
 
\section{Introduction}

Nearly all motile animals rely on proprioceptive feedback 
for the control of the body \cite{tuthill2018proprioception}.
An example is the proprioceptive measurement of limb angles, 
which has a resolution of about 1$^\circ$ for humans and of 
roughly 10$^\circ$ for flies \cite{tuthill2018proprioception}.
For humans, the deprivation of the capability to sense limb 
postures via muscle tensions leads to the complete inability
to perform coordinated movements, viz to immobility
\cite{mcneill2010iw,tuthill2018proprioception}. 
Without the internal sensory feedback from the body, viz propriosensation, humans can activate muscles
only individually, but not perform coordinated physical 
actions, such as sitting or walking. 

\begin{figure}[t]
\centering
\includegraphics[width=0.8\textwidth]{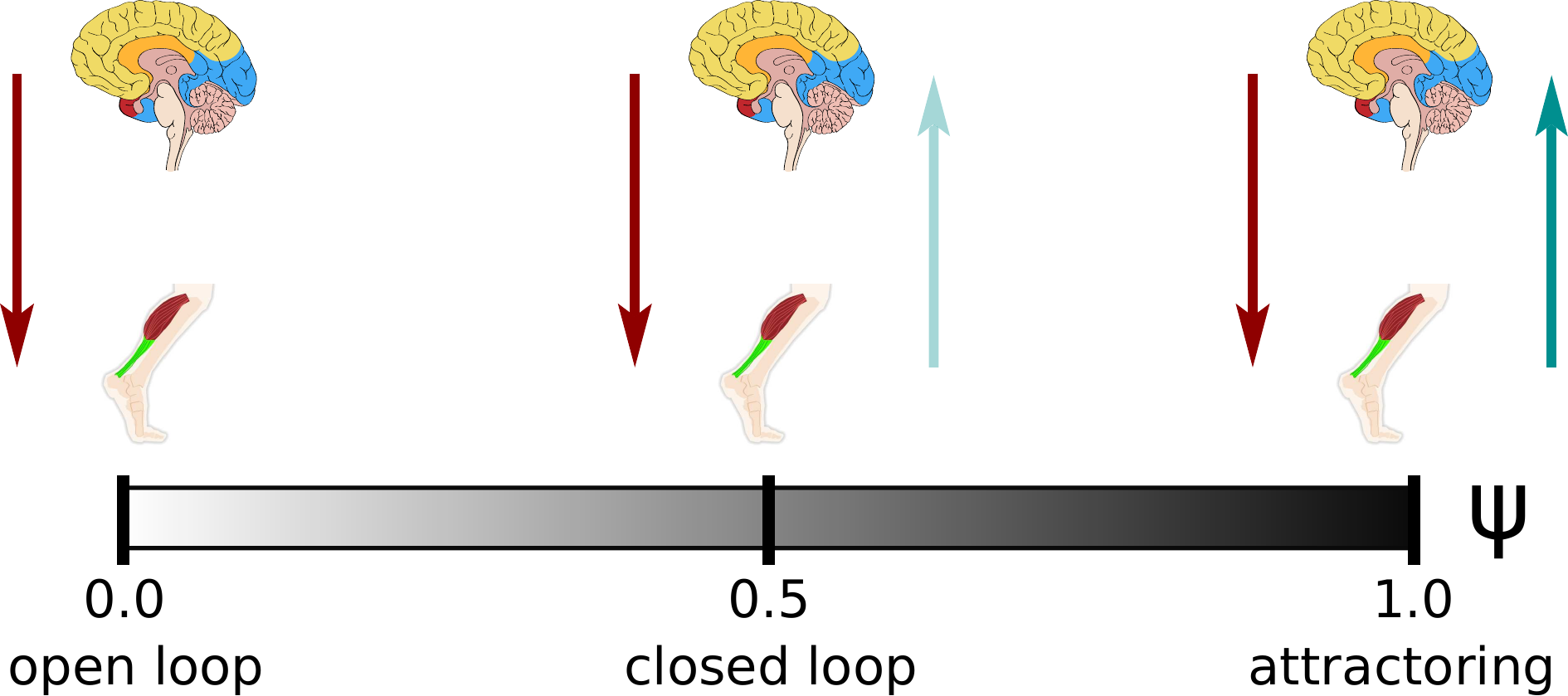}
\caption{{\bf Control schemes.} 
As a function of the mixing angle $\Psi\!\in\![0,1]$, the 
transition between open-loop, closed-loop, and self-organized 
attractoring. Shown are motor commands (red down-arrows) and 
proprioception (cyan up-arrows), with the control units (top) 
receiving proprioceptive signals from the actuators (bottom). 
Motor commands are generated independently of the
state of the body for open-loop control ($\Psi\!=\!0$, left),
being modulated, but not driven, for closed-loop control 
($0\!<\!\Psi\!<\!1$, middle). Attractoring is present when the
generation of motor commands is to 100\% dependent on 
proprioception ($\Psi\!=\!1$, right). In this limit, there is no 
locomotion when the feedback loop is cut. Animals tend to have
high mixing angles $\Psi\!\sim\!1$, which implies that
the attractoring limit $\Psi\!\to\!1$ may be considered
as a default for modeling approaches.
}
\label{fig_controlTypesIllustration}
\end{figure}

From a general perspective, locomotion may be induced on 
three levels, by open-loop, closed-loop, and self-organized
control. See Fig.\,\ref{fig_controlTypesIllustration}.
For the first, actuators and limbs do not signal back 
the result of the control signals. Open loop top-down 
control principles are important in predictable 
situations, e.g.\ when stick insects move on flat surfaces
\cite{bassler1998pattern}, and when the time scale of
locomotion is faster than the delay time inherent
in the proprioceptive feedback loop \cite{tuthill2018proprioception}.
For the second case, closed-loop control, feedback signals
from the actuators modulate the functioning of the circuits 
generating the motor commands. For periodic movements, such 
as slow motion on rough terrain \cite{bassler1998pattern}, this 
implies that the parameters of a central pattern generator
(CPG) are continuously readjusted. In the third case,
self-organized control, motor commands allowing locomotion 
are not generated
at all in the absence of proprioceptive feedback, which implies
that agents are immobile when deprived of propriosensation.
Compliant locomotion is generated in terms of
self-stabilizing attractors that form in the 
sensorimotor loop, a route to locomotion denoted here 
as `attractoring', or 'self-organized attractoring'.

Open-loop control and attractoring are limiting
cases of closed-loop control, as
illustrated in Fig.\,\ref{fig_controlTypesIllustration}.
Parameterizing the relative impact of the proprioceptive
feedback on the motor-command-generating circuits by an
abstract mixing angle $\Psi\!\in\![0,1]$, closed-loop control 
is present whenever $0\!<\!\Psi\!<\!1$, with open-loop 
control and attractoring corresponding respectively to 
the limits $\Psi\!\to\!0$ and $\Psi\!\to\!1$. For
small mixing angles, say $\Psi\!\approx\!0.1$, the
motor signals are only mildly readjusted by sensory 
feedback signaling. Motor commands are determined on 
the other side to a large extent by the sensory feedback
when the mixing angle is large, e.g.\ for $\Psi\!\approx\!0.9$.
When deprived of sensory feedback, locomotion will still 
be functional at large mixing angles, albeit at the 
expense of a strongly reduced quality. 

Control systems characterized by small and large mixing 
angles can be approximated to first order respectively
by open-loop control and attractoring. It has been 
argued \cite{schilling2019decentralized}, that the 
neuronal circuits controlling insect locomotion cover 
the complete range of mixing angles $\Psi$, with 
the effective mixing degree being a function of walking 
speed and environmental factors. Given that slow-moving
animals tend to operate at high mixing angles 
\cite{tuthill2018proprioception}, close to the regime
of self-organized attractoring, the limiting case 
$\Psi\!\to\!1$ deserves attention. It is also
interesting to note, that most state-of-the-art 
machine learning algorithms for motor control 
tasks operate implicitly in the attractoring limit.
Generically the effect of the motor commands on
the physical system is measured, within the current 
machine learning frameworks
\cite{lillicrap2015continuous,duan2016benchmarking}, 
with the respective sensor reading driving the 
commanding network.

\begin{figure}[t]
\centering
\includegraphics[width=.8\textwidth]{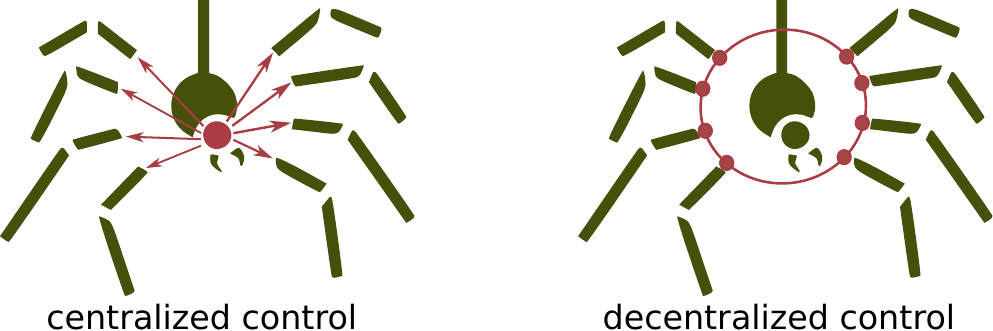}
\caption{{\bf Centralized vs.\ emerging limb coordination.} 
For centralized control (left) a CPG outputs motor commands
to the individual actuators. Limb coordination is the 
responsibility of the CPG. For decentralized control (right),
motor commands are generated locally. Limb coordination 
emerges from the interaction of the individual actuators,
either via neuronal inter-actuator information exchange 
or through the physical response of the body, viz via embodiment.
In real-world applications, both for animals and robots, 
a mixture of centralized and decentralized
control is observed.
}
\label{fig_deCentralizedControl}
\end{figure}

\subsection{Modeling animal and robotic locomotion}

Locomotion is about coordinated movement
\cite{pfeifer2001understanding}. A classical 
route to achieve coordinated activation of actuators
is the use of central pattern generators 
\cite{ijspeert2008central}, which are well suited to
produce regular muscle contractions, like breathing 
\cite{smith2013brainstem} and gaits \cite{marder2001central}, 
possibly also for biped locomotion, viz for human 
walking \cite{minassian2017human}. The influence of
feedback from internal sensors onto the CPG can be
included in various fashions \cite{aoi2017adaptive}, 
f.i.\ through chaos control \cite{steingrube2010self}.
Going one step further, an interesting question is
whether higher cognitive functions may evolve from
locomotion-controlling frameworks \cite{schilling2017reacog,koglin2019when}.

As an alternative to CPGs, actuators may be controlled
by local circuits. Varying the centralization degree
\cite{neveln2019information}, a continuum of control
schemes interpolating between the two endpoints, fully
centralized and distributed control, is attained.
See Fig.\,\ref{fig_deCentralizedControl}. An example of 
decentralized control is the local phase 
oscillator \cite{owaki2013simple},
\begin{equation}
\dot\phi_i = \omega - \sigma N_i\cos(\phi_i)\,,
\label{localOscillotor}
\end{equation}
where $\phi_i$ is a phase that is specific to the $i$th
actuator, $\omega$ the natural frequency~\cite{gros2015complex}, 
$N_i$ a locally measured force, like the ground-reaction
force, and $\sigma$ the self-coupling constant.
It has been shown \cite{owaki2013simple,owaki2017quadruped},
that robust locomotion arises for quadruped robots for 
which the motor commands for the legs $i=1,2,3,4$ are
generated locally by oscillators of type (\ref{localOscillotor}).
Various gaits, in particular walking, trotting, and 
galloping \cite{owaki2017quadruped}, are induced solely by 
mechanical inter-limb interactions. Physically, the
measured ground force $N_i$ allows the leg to enter 
the swing phase only once the load on the leg has 
decreased sufficiently, which happens when other legs 
start to carry a fair share of the weight by touching 
the ground. Similar results for robots driven by 
actuator-specific oscillators were
found for hexapods \cite{ambe2018simple}.

The emergence of inter-actuator coordination via the mechanics
of the body can be studied also in the context of wheeled
robots \cite{kubandt2019embodied}. For simulated trains
of five two-wheel cars, for which the wheels are controlled 
individually by a one-neuron attractoring scheme, it has been found 
that the $10\!=\!5\cdot2$ wheels coordinate their rotational 
frequencies to produce highly compliant behavior. The train of 
cars is able move in a snake-like fashion, to turn autonomously 
when climbing a slope, to accelerate downhill and to interact
non-trivially with the environment, f.i.\ by pushing around a
movable box \cite{kubandt2019embodied}.

Part of the computational effort that is needed to generate
robust locomotive patterns  may be carried out
by the body and its elasto-mechanical constituents
\cite{aguilar2016review}. When this happens, one speaks 
of `embodiment' \cite{pfeifer2007self}. Examples are
passive walkers \cite{collins2005efficient}, dead rainbow
trouts swimming upstream in vortex wakes \cite{beal2006passive}, 
and the self-organized inter-leg communication via the mechanical
properties of the body \cite{owaki2013simple,owaki2017quadruped}, as
discussed further above in conjunction with Eq.~(\ref{localOscillotor}).
Embodiment can be viewed as an instance of morphological 
computation \cite{muller2017morphological,ghazi2017morphological},
which stresses the role that bodies, in particular soft bodies 
like octopus arms \cite{guglielmino2012application}, have
for compliant movements \cite{hauser2011towards}. Suitable 
approaches for the selection of the control circuits of 
embodied agents are, besides other, evolutionary algorithms 
\cite{floreano2000evolutionary,agmon2014evolution} and the 
principle of guided self-organization 
\cite{prokopenko2009guided,gros2014generating},
where the latter may be implemented in terms of a stochastic 
attractor selection mechanism \cite{nurzaman2014guided}.

Complementary to efforts dedicated to develop theoretical 
frameworks \cite{roth2014comparative}, the focus of the 
present overview, a substantial number of studies have 
been dedicated to the modeling of animal locomotion 
on a detailed biological level
\cite{orlovskiui1999neuronal,ayali2015comparative}.
Starting from central pattern generators
\cite{marder2001central,arshavsky2016central}, it
has been realized that observed walking patterns 
are at times difficult to classify into distinct gait 
classes \cite{schilling2019decentralized}. Instead,
movement patterns seem to form a continuous
two dimensional manifold \cite{deangelis2019manifold}.
Complete models may contain a substantial number of 
differential equations \cite{ayali2015comparative},
which is typically of the order of 52 
\cite{azevedo2019size} to 164 
\cite{schilling2019decentralized} units per limb.
A human leg, as a comparison, is innervated by over 
150,000 motor neurons \cite{kernell2006motoneurone}.

A recurring notion emerging from experimental studies
regards the key importance of sensorimotor interactions 
for locomotive behavior 
\cite{cohen2014nematode,frost2015sensorimotor,klein2012physical}.
Formally one defines the `sensorimotor loop' as a
dynamical system that is defined within the combined 
state space of environmental degrees of freedom, body, 
actuator, and sensory readings \cite{sandor2015sensorimotor}.
Within this state space, dynamical attractors may form,
with fixpoints corresponding to inactivity and limit-cycles
to rhythmic behavior \cite{martin2016closed}. Attractors
in the sensorimotor loop correspond to motor primitives
that can be used as the basis of more complex behavior.
Secondary control, like `kick control' schemes, enable 
then an overarching control unit, e.g.\ the brain, to 
generate sequences of locomotive states in terms of 
motor primitives \cite{sandor2018kick}. Kick 
control can be viewed in this context as an instance of 
a higher-level control mechanism that exploits the reduction 
in control complexity provided by embodied robots 
\cite{montufar2015theory}.

A series of theoretical concepts aim to formalize the
role of the sensorimotor loop for locomotion, in
particular for embodied agents. One possibility is
to maximize the predictive information generated 
within the sensorimotor loop \cite{zahedi2010higher},
other proposals elucidate the role of short-term
synaptic plasticity \cite{martin2016closed,toutounji2014behavior}
and differential extrinsic plasticity 
\cite{der2015novel,pinneri2018systematic}. 

Here, we aim to provide a compact overview of dynamical 
systems approaches of robotic locomotion in the 
attractoring limit, with a focus on basic
concepts. We will stress that attractoring is, 
despite its relevance in particular for animal 
locomotion, a hitherto comparatively unexplored 
area of robotic control. In Sect.\,\ref{sect_DC_principle} 
we review a basic generative mechanism for attractoring,
the `Donkey \@ Carrot' (DC) principle. 
Sect.\,\ref{sect_all_robots} then illustrates
that for a given generative mechanism, here the
DC principle, a range of options of how to 
implement the algorithm for simulated and real-world 
robots exist (see Fig.~\ref{fig:robots}). 

\section{The Donkey \& Carrot principle 
         for self-organized actuators\label{sect_DC_principle}}

A well-known metaphor concerns a donkey and a
carrot. A boy riding a donkey uses a pole to hold
a carrot in front of the animal, which locomotes in
an attempt to reach the carrot, however without
ever attaining the goal. When generalized to the
sensorimotor loop, this principle, the Donkey \&
Carrot (DC) principle, leads to self-sustained
locomotion in terms of limit-cycle attractors.

The starting point of the DC algorithm is the
actual state $s^{(\mathrm{a})}$ of the actuator, which is 
the state given by a real-time measurement. This
state, $s^{(\mathrm{a})}$, is transformed into an input 
signal, denoted $y^{(\mathrm{s})}$, which has a magnitude 
and a range that is suitable for the local
controlling circuit. Driven by this input, $y^{(\mathrm{s})}$,
the local controller produces an output $y$.
The output is then transformed into a target state
$s^{(\mathrm{t})}$ for the actuator, viz into the state the 
actuator is supposed to reach. For this purpose a motor signal 
proportional to the difference $s^{(\mathrm{t})}-s^{(\mathrm{a})}$ 
is generated. This procedure is repeated at every cycle 
of the control loop, each time with the newly measured 
actual state $s^{(\mathrm{a})}$.

For the self-organized DC actuator discussed here,
the target state $s^{(\mathrm{t})}$ will be reached only for 
fixpoints, viz for non-moving solutions, but not for 
locomotive states. The mechanism is that $s^{(\mathrm{t})}$ changes,
continuously, whenever the actual position $s^{(\mathrm{a})}$
changes in response to the motor signal and the
environmental feedback. In contrast to a stiff actuator 
with a close-to-perfect and instantaneous response, 
the compliance of the self-organized DC actuator 
allows for the interaction between multiple limbs 
and between the body and the environment, which can 
directly influence each other's dynamics. 
In this way, the feedback may also result in a 
self-organized coordination of the different joints 
and an autonomous reaction to a changing environment.

This feedback principle, which
has been shown to generate robust and highly compliant 
locomotion \cite{martin2016closed}, is universal in 
the sense that it can be applied to a wide range of 
actuator types, including weights moving along a 
rod \cite{sandor2015sensorimotor} and standard 
wheeled robots \cite{sandor2018kick}.

\subsection{A one-neuron DC actuator}
\label{sect_DC_actuator}

In its simplest implementation, the DC controller
employs a single rate-encoding neuron. We define 
with $x$ the membrane potential of
the neuron. A standard leaky integrator,
\begin{equation}
\tau_x\dot{x} = w\,y^{(\mathrm{s})} - x
\label{dot_x}
\end{equation}
is used for the evolution rule. In (\ref{dot_x})
the time scale of the membrane potential is given 
by $\tau_x$, with the synaptic weight $w>0$ 
coupling the proprioceptual input $y^{(\mathrm{s})}$ to 
the controlling neuron. Note that (\ref{dot_x}) 
can be viewed also as a low-pass filter 
\cite{chen2012smooth}. The proprioceptual activity 
$y^{(\mathrm{s})}$ is normalized, $y^{(\mathrm{s})}\in[0,1]$, 
which is attained by
\begin{equation}
y^{(\mathrm{s})}= \frac{s^{(\mathrm{a})}-s^\mathrm{(min)}}
{s^\mathrm{(max)}-s^\mathrm{(min)}},
\qquad\quad
s^{(\mathrm{a})}\in[s^\mathrm{(min)},s^\mathrm{(max)}]\,,
\label{y_s}
\end{equation}
where $s^{\mathrm{(min)}}$, $s^{(\mathrm{a})}$, and $s^{\mathrm{(max)}}$ 
denote respectively the minimal, the actual, and
the maximal values for the state of the actuator. 
This expression for
$y^{(\mathrm{s})}$ is valid whenever $s^\mathrm{(min)}<s^\mathrm{(max)}$.
For the case of a wheel, which is characterized by an
angle $\varphi\in[0,2\pi]$, one takes
$y^{(\mathrm{s})}\to\cos(\varphi)$, which implies in this case
that $y^{(\mathrm{s})}\in[-1,1]$~\cite{sandor2018kick}.

A rate encoding neuron is defined by the transfer
function $y(x)$, for which we consider a sigmoidal,
\begin{equation}
y(x) = \frac{1}{1+e^{a(b-x)}},
\qquad\quad y\in[0,1]\,,
\label{y_x}
\end{equation}
which is parametrized by a gain $a>0$ and a threshold $b$. 
The neural activity generates the target position 
$s^{(\mathrm{t})}$ for the actuator, here via
\begin{equation}
s^{(\mathrm{t})} = (1-y(x))s^{\mathrm{(min)}}+y(x)\,s^{\mathrm{(max)}}\,,
\qquad\quad s^{(\mathrm{t})}\in[s^{\mathrm{(min)}},s^{\mathrm{(max)}}]\,,
\label{s_t}
\end{equation}
which represents a linear mapping of $y(x)$ to the 
allowed range $[s^{\mathrm{(min)}},s^{\mathrm{(max)}}]$
of the state of the actuator. 
As the differential equations (\ref{dot_x}) 
explicitly depend on the proprioceptive input signals, 
they have to be solved online by the local 
computer/microcontroller of the robot using some 
numerical algorithm. In the examples below we could 
achieve time steps as small as 40-50 ms 
that allows for a numerically stable solution 
even when combined with the Euler method.

For robots equipped with stepper motors, the target
position $s^{(\mathrm{t})}$ can be used directly.
Otherwise, a motor signal $F_k$ corresponding to 
the force, or to the torque, for the case of wheels, 
needs to be generated. For the simplest approach,
\begin{equation}
F_k = k\left(s^{(\mathrm{t})} - s^{(\mathrm{a})}\right),
\qquad\quad
F_\gamma = -\gamma \frac{\mathrm{d}}{\mathrm{d}t}
\left(s^{(\mathrm{t})} - \ s^{(\mathrm{a})}\right)\,,
\label{F_motor}
\end{equation}
the dynamics generated by a spring with a spring constant
$k$ is simulated.  Commercially available motors will 
be controlled in practice by a PD controller, which implies 
that a damping term $F_\gamma$, as defined in 
(\ref{F_motor}), will be active. For simulated
robots, one can select $k$ and the damping coefficient
$\gamma$ by hand. A constant target state $s^{(\mathrm{t})}$ 
is approached smoothly under (\ref{F_motor}).

For not-too-high spring constants $k$, the actuator 
responds softly to the control signal, while being strongly
influenced by the feedback of the environment via 
proprioception. 
The actuator is therefore compliant by the control 
\cite{sandor2015sensorimotor}, which does not exclude 
additional compliance due to the structure of the body, 
or to soft constituents \cite{pfeifer2007self}. 
In contrast to classical target following control, 
most of the time, the target state $s^{(\mathrm{t})}$ is only 
followed with a delay, but not reached, according to the 
Donkey \& Carrot principle. The resulting dynamics, which 
can be modulated by changing the parameters of the 
DC controller is thus self-organized, through the 
continuous interaction of brain, body, and environment.

\begin{figure}[t]
\centering
\includegraphics[width=1.0\textwidth]{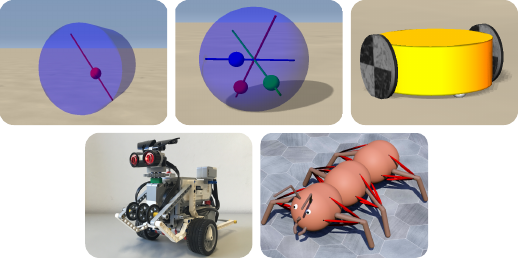}
\caption{{\bf Robots powered by local DC controllers.}
Shown are examples of investigated simulated and real-world robots.
\textit{Top row:} Simulated robots from the 
LPZRobots simulations environment \cite{der2012playful}, 
which include barrel-, sphere-, and two-wheeled robots.
\textit{Bottom row}: A two-wheeled Lego Mindstorms 
robot (left) and a muscle-driven hexapod simulated
within the simulation platform Webots \cite{Webots}. 
Embodied coordination of the 24 muscles induces locomotion
via a stable limit cycle
(\href{https://doi.org/10.6084/m9.figshare.23703399}{video}).
	}
\label{fig:robots}
\end{figure}

\subsection{Self-organized embodiment}

The here presented dynamical-systems framework 
enables the design and construction of fully embodied 
robots. Furthermore, it also allows for a stringent 
definition of self-organized embodiment. The term 
embodiment is used, in particular
in the context of cognitive robotics, whenever the 
behavior of an active agent is not simply the 
outcome of its internal motivation, but when it 
results from the ongoing interaction with the 
environment \cite{pfeifer2007self}.

Here, we reserve the term `self-organized embodiment' for
emergent behavior that cannot be reproduced by isolated 
controllers and actuators, that is by a robot that is 
separated from the environment. Within the terminology 
of dynamical systems, self-organized embodied dynamics 
is characterized by the presence of attractors that cease 
to exist when the subsystem of the controller is isolated. 
This is the case for the DC controller described by
(\ref{dot_x}), (\ref{y_s}) and (\ref{s_t}). 

In order to see why, consider the instantaneous
approximation $s^{(\mathrm{a})}\equiv s^{(\mathrm{t})}$
which implies that the environment has no time
to react, and hence no influence.
Within this assumption of instantaneous actuators,
one has an open-loop control scheme that reduces to 
the autapse condition $y^{(\mathrm{s})}=y(x)$ in (\ref{y_x}). 
The environment is then short-circuited and left out 
of the control process, which results in a stiff 
controller. From (\ref{dot_x}) it follows that 
$x^*=wy(x^*)$, where $x^*$ is a fixpoint of the 
membrane potential. Locomotive limit cycles
are hence absent.

Our definition of self-organized embodiment distinguishes
self-organized controllers, like the DC controller,
from embodied approaches that rely on local pace-making
circuits \cite{owaki2017quadruped,ambe2018simple}. Robots
that are powered by limbs that are autonomously active even
in the absence of feedback from the environment represent 
in this perspective a different type of embodiment.

\begin{figure}[t]
\centering
\includegraphics[width=1.00\textwidth]{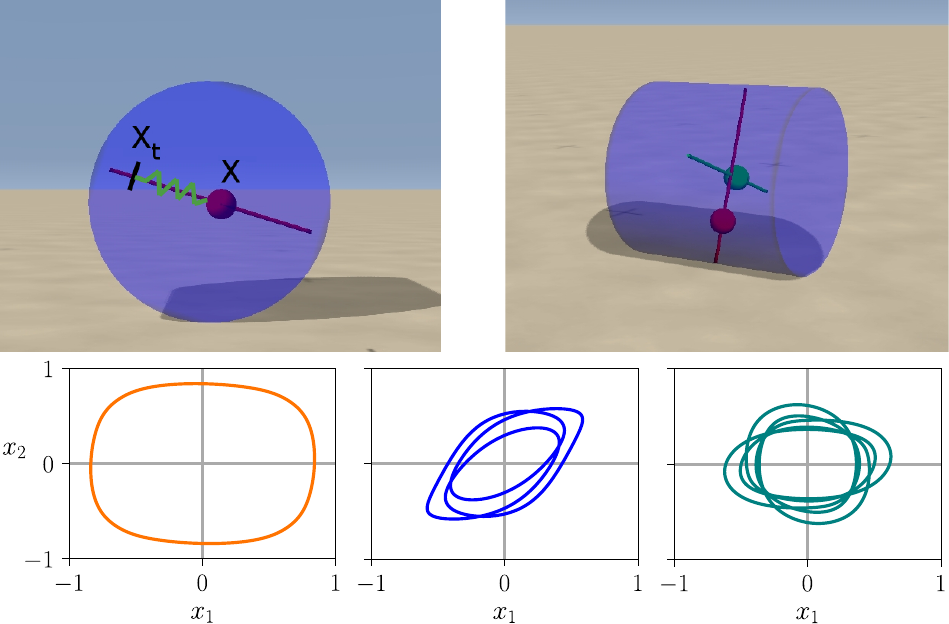}
\caption{{\bf Self-organized barrel robot.}
Barrel-shaped robots simulated with
the LPZRobots simulations environment \cite{der2012playful}.
\textit{Top-right:} With two independent actuators (green/red), 
each being composed of a weight moving on a rod within the body 
of the robot. In this case 
(\ref{dot_x}) has been used \cite{sandor2015sensorimotor}.
\textit{Top-left:} Illustration of the actual and the
target position $x\equiv s^{(a)}$ and 
$x_t\equiv s^{(t)}$. A simulated spring (green)
pulls the weight towards the target position. Only one of 
the two rods is shown.
\textit{Bottom:} The two-rod barrel rolls with 1:1, 1:3 and 1:5 
frequency locking upon changing parameters (left to right), 
needing respectively 1/3/5 revolutions in the $x_1-x_2$ plane
for a closed orbit. Here $x_1$ and $x_2$ are the positions
of the two weights
(\href{https://doi.org/10.6084/m9.figshare.11809068.v1}
{video}).
}
\label{fig_barrel_robot}
\end{figure}

\section{Self-organized embodied simulated 
         and real-world robots\label{sect_all_robots}}

The framework presented in Sect.\,\ref{sect_DC_actuator},
with locomotion that results from self-organized embodiment,
is quite generic. Specific implementations are possible
for a wide range of distinct morphologies, which
include barrel- and sphere-shaped robots, wheeled
robots, train of wheeled cars, and legged robots 
such as hexapods. 

\subsection{Barrel robots\label{sect_barrel}}

In Fig.\,\ref{fig_barrel_robot} we show
a barrel-shaped robot that is driven by two
independent actuators composed each of a weight
moving along a rod. One finds a surprisingly rich
phase diagram in terms of the internal
parameters \cite{sandor2015sensorimotor}, such 
as the gain~$a$, entering (\ref{y_x}), and 
the adaption rate~$\epsilon\sim 1/\tau_b$, 
where $\tau_b$ determines the time scale of 
internal adaptation of the threshold $b$ according to:
\begin{equation}
\tau_b \dot b = y(x)-\frac{1}{2}\,.
\label{dot_b}
\end{equation}
For the barrel robot we set $x=y^{(\mathrm{s})}$ and $w=1$, 
the instantaneous limit of (\ref{dot_x}), adapting 
instead with (\ref{dot_b}) the threshold $b=b(t)$, 
which ensures that the fixpoint $y^*=1/2$ is unstable. 
The Donkey \& Carrot framework remains otherwise untouched.
The two actuators coordinate their movements spontaneously 
via the mechanics of the body, as one can observe in 
the video included in Fig.\,\ref{fig_barrel_robot}, 
with phase matching occurring in 1:1, 1:3 or 1:5 modes, 
in terms of the number of revolutions of the internal 
weights corresponding to one rotation of the barrel,
as parameters are varied.

\begin{figure}[t]
\centering
\includegraphics[width=1.0\textwidth]{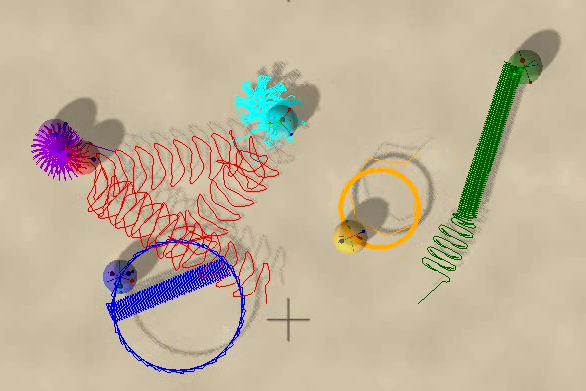}
\caption{{\bf Attractoring for self-organized sphere robots.}
Shown are six simulated sphere robots (top view) that are powered by 
the same self-organizing principle, but with different parameter
settings, $(w,z)$. Each robot is actuated by three weights that
move in the inside along perpendicular rods, see
Fig.\,\ref{fig:robots} for a blowup. For each
weight, a single neuron produces a target position, with
the input given by the actual position of the weight
and inhibitory feedback from the two other neurons.
\cite{martin2016closed}. The lines retrace the past 
trajectories in the plane of locomotion.
Interaction between the robots leads to changes 
in the attracting state, which corresponds either 
to a transition to a different mode (blue and green robots),
or to a different phase or direction (red, cyan, 
and magenta robots).  This type of behavior is 
called here `attractoring'
(\href{https://figshare.com/articles/Locomoting_attrractors_of_self-organized_sphere_robots/7874606}
{video}).
}
\label{fig_sphere_robot}
\end{figure}

\subsection{Sphere robots\label{sect_sphere}}

The sphere robot is driven by weights moving
along three perpendicular rods. In this case,
Eq.~(\ref{dot_x}) incorporates a direct inhibitory 
coupling (proportional to the weight $z$) 
between the three actuators:
\begin{equation}
\tau_x\dot{x}_i = w\,y_i^{(\mathrm{s})}-z\sum_{k\ne i} y_ku_k\varphi_k- x_i\,.
\qquad\quad z>0\,,
\label{dot_x_sphere}
\end{equation}
The time-dependent parameters $u_i=u_i(t)$ and 
$\varphi_i=\varphi_i(t)$ modulate
the synaptic strength temporally, a phenomenon denoted
short-term synaptic plasticity (STSP) \cite{zucker2002short}.
For the STSP, which depends exclusively on the activity 
level of the presynaptic neuron, a modified version 
\cite{sandor2017complex} of the original Tsodyks and Markram
model was taken \cite{tsodyks1997neural}. One finds, as 
shown in the video enclosed with Fig.\,\ref{fig_sphere_robot}, 
a rich repertoire of limit-cycle dynamics that leads to 
various gaits for forward and circle-shaped locomotion
\cite{martin2016closed}. When two sphere robots collide, 
they are able to kick each other into alternative 
attractors, which may be either of a distinct type or 
oriented differently with respect to the
direction of propagation.

\begin{figure}[t]
\centering
\includegraphics[height=0.38\textwidth]{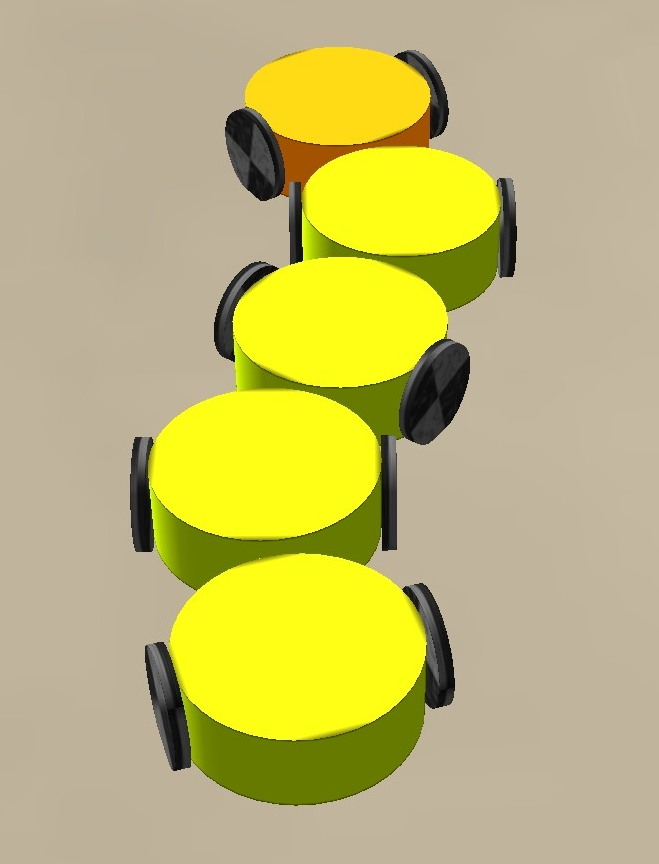}
\includegraphics[height=0.38\textwidth]{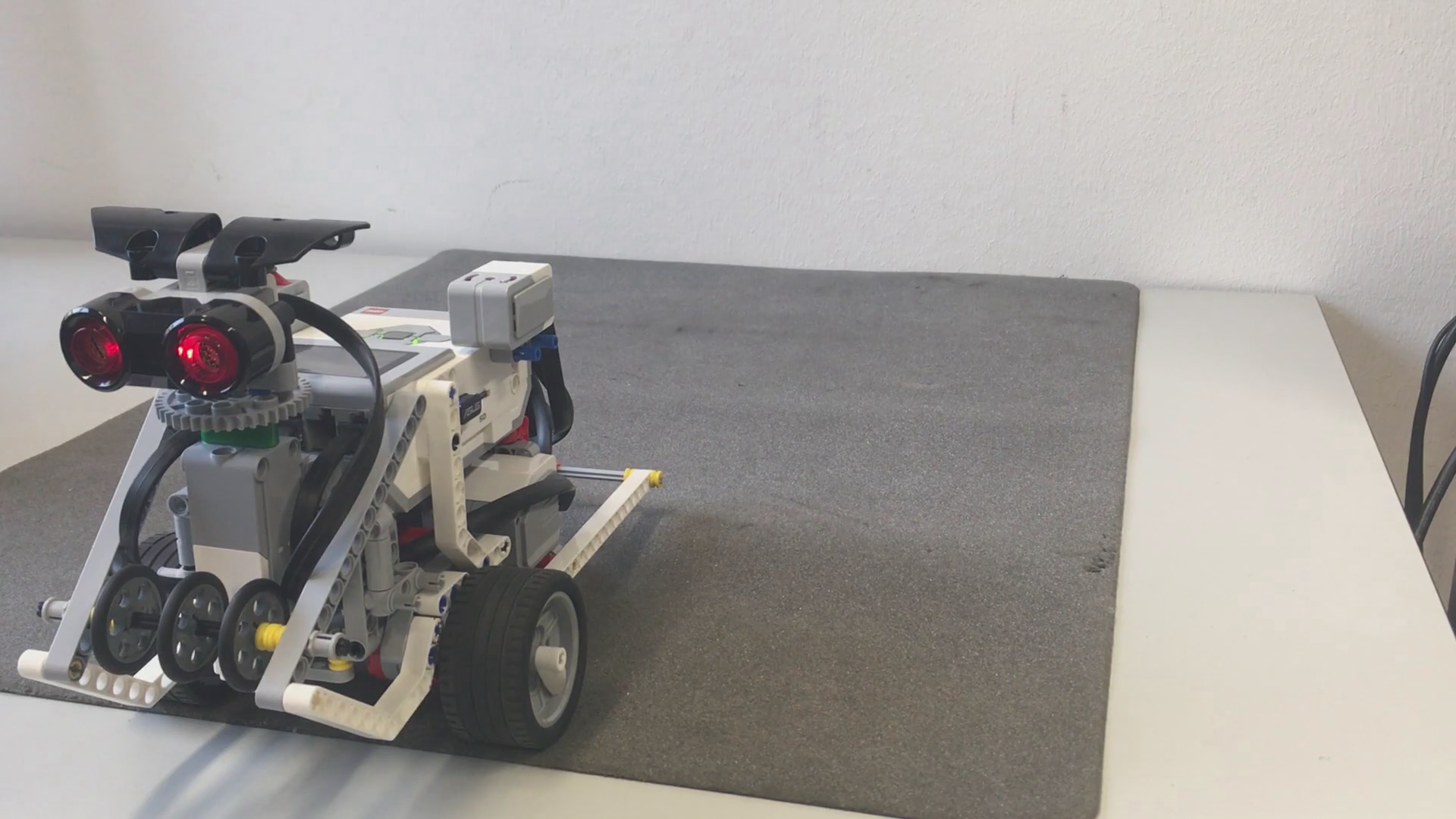}
\caption{{\bf Self-organized wheeled robot.}
\textit{Left:} A simulated train of cars, with each of 
the ten wheels being controlled by a one-neuron DC 
controller \cite{kubandt2019embodied}. The cars are 
coupled by passive hinge joints. Inter-wheel communications 
occurs solely via the mechanical response of
the body. Coordinated locomotion that explores the
given environment in a non-trivial manner emerges
(\href{https://figshare.com/articles/carChain_slope_pushing_mp4/7643123/1}
{video}).
\textit{Right:} 
A Lego Mindstorms robot. Both wheels are actuated 
independently by the Donkey \& Carrot actuator,
see Sect.\,\ref{sect_DC_actuator}, here with two neurons
per wheel \cite{sandor2018kick}. Due to time-reversal 
symmetry attractors emerge in forward/backward pairs. 
Interacting with the environment, a wall bounce, the robot 
is kicked from the forward into the backward attractor. 
The robot does not sense the wall, it only knows about 
the angle of its wheels
(\href{https://figshare.com/articles/Autonomous_direction_reversal_of_an_embodied_wheeled_robot/7880393}
{video}).
}
\label{fig_train_car_Lego_robot}
\end{figure}

\subsection{Wheeled robots\label{sect_wheel}}

Wheels turn continuously, which implies that
there is no minimal or maximal value for the 
state of the actuator. One then substitutes
$y^{(\mathrm{s})}=\cos(\varphi)$ for (\ref{y_s}), which implies
that $y^{(\mathrm{s})}\in[-1,1]$. The DC controller remains
otherwise the same. For the Lego Mindstorms robot
presented in Fig.~\ref{fig:robots} 
and Fig.\,\ref{fig_train_car_Lego_robot}, two
neurons per wheel have been used, with the second
neuron taking $y^{(\mathrm{s})}=\sin(\varphi)$ as its driving 
input. This configuration can be interpreted as 
two perpendicular simulated transmission rods \cite{sandor2018kick},
in the style of the transmission rod of
classical steam engines. The Lego robot shows
chaotic and limit-cycle behavior, with the latter
being twofold degenerate. Time-reversal symmetry
demands that there is a limit cycle corresponding to
backward motion whenever there is one for moving
forward, and vice versa. The robot may hence be
kicked from the forward into the backward attractor
when interacting with the environment, as it
occurs in the video included with
Fig.\,\ref{fig_train_car_Lego_robot}. 
This emergent behavior is remarkable in the view that 
no such specific function was implemented explicitly 
in contrast to the classical robotics approaches.
Alternatively, one can use a top-down kick control signal to
induce motion reversal or changing the direction of 
locomotion by turning around the vertical axis of the robot 
\cite{sandor2018kick}.

Also shown in Fig.\,\ref{fig_train_car_Lego_robot}
is a simulated train of passively coupled two-wheel
cars~\cite{kubandt2019embodied}, compare 
Fig.~\ref{fig:robots}. Single wheels are 
actuated by a one-neuron DC controller, with 
inter-wheel coordination happening solely due 
to the mechanical response of the robot components.
Snake-like locomotion, autonomous direction
reversal and non-trivial interaction with the
environment, like pushing around a movable box,
emerges spontaneously. A link to a video is
included in the caption of Fig.\,\ref{fig_train_car_Lego_robot}. 

\subsection{Muscle driven hexapod}

Attractoring can be generalized to muscle
driven animats, as illustrated in
Fig.\,\ref{fig:robots}. Using the physics 
simulation platform Webots \cite{Webots},
we constructed a hexapod with four muscles
per leg, each driven by an independent attractoring
feedback loop \cite{fischer2023neural}. Stable 
locomotion emerges here via `force coupling' \cite{fischer2023neural}, a controller scheme 
for which the firing of a single neuron 
influences the contraction of multiple muscles 
but not directly the activity of other neurons. 
Direct couplings between the 24 local control 
loops are absent, which implies that the 
coordination between the legs is 100\% embodied.

\section{Conclusion}

Robots are used for large varieties of purposes,
which range from industrial applications to the 
modeling of animal behavior. From the perspective
of living machines, it is in this context important 
to explore routes to locomotion independently of 
whether they provide an immediate improvement over 
existing control schemes. A particularly interesting 
framework, self-organized embodiment, suggests
a modular approach, consisting of limbs that are
locally controlled, with interlimb coordination
remaining the task of either morphological computation,
via the body, or of decentralized control circuits.
Self-organized embodiment has the potential to
reduce the complexity of the control task by
making use, e.g.\ via kick control, of the 
set of motor primitives generated autonomously
within the sensorimotor loop. The present framework 
allows us to carry out a full mapping of the parameter 
space not only finding some optimal values but 
also understanding the role of each parameter. 
For more complex applications, one could 
rely on optimization algorithms to find the best 
parameters for some specific task. Here we presented
a review of the state of the field.

\begin{credits}
\subsubsection{\ackname} The work of BS was supported by the grant of the Romanian Ministry of Research, Innovation and Digitization, CNCS - UEFISCDI, project number PN-III-P1-1.1-PD-2019-0742 within PNCDI III, and SRG-UBB 32993/23.06.2023 within UBB Starting Research Grants of the Babe\textcommabelow{s}-Bolyai University.
\subsubsection{\discintname}
The authors have no competing interests to declare that are
relevant to the content of this article. 
\end{credits}

%
%
%
\bibliographystyle{unsrt}
\bibliography{attractoring_robots}


\end{document}